\documentclass[prl,reprint,superscriptaddress]{revtex4-1}
\usepackage{soul}
\usepackage{color}
\usepackage{graphicx}
\usepackage{amsmath}
\usepackage{array}

\usepackage{float}
\usepackage{lineno}
\usepackage{ulem}
\usepackage{siunitx}
\usepackage[bottom]{footmisc}
\usepackage{ragged2e}
\usepackage[dvipsnames]{xcolor}
\definecolor{apsblue}{HTML}{2E3092}
\usepackage[colorlinks=true,
            linkcolor=apsblue,
            citecolor=apsblue,
            urlcolor=apsblue]{hyperref}
\begin{document}

\title{Probing large mass-splitting inelastic Dark Matter with RES-NOVA}


\def\there{Work}
\def\here{Home}

\def\udel{Department of Physics and Astronomy, University of Delaware, Newark, DE 19716, USA}
\def\udelbartol{Department of Physics and Astronomy, University of Delaware and the Bartol Research Institute, Newark, DE 19716, USA}

\author{D.~Alloni}
\affiliation{Laboratorio Energia Nucleare Applicata, Via Aselli 41, I-27100 Pavia, Italy}

\author{G.~Benato}
\affiliation{Gran Sasso Science Institute, Viale F. Crispi 7, I-67100 L’Aquila, Italy}
\affiliation{INFN Laboratori Nazionali del Gran Sasso, Via G. Acitelli 22, I-67100 Assergi, Italy}

\author{P.~Carniti}
\affiliation{Dipartimento di Fisica, Università di Milano - Bicocca, Piazza della Scienza 3, I-20126 Milano, Italy}
\affiliation{INFN Sezione di Milano - Bicocca, Piazza della Scienza 3, I-20126 Milano, Italy}

\author{M.~Cataldo}
\affiliation{Dipartimento di Fisica, Università di Milano - Bicocca, Piazza della Scienza 3, I-20126 Milano, Italy}
\affiliation{INFN Sezione di Milano - Bicocca, Piazza della Scienza 3, I-20126 Milano, Italy}

\author{L.~Chen}
\affiliation{Shanghai Institute of Ceramics, CAS, 1295 Dingxi Road, Shanghai 200050, P.R. China}

\author{M.~Clemenza}
\affiliation{Dipartimento di Fisica, Università di Milano - Bicocca, Piazza della Scienza 3, I-20126 Milano, Italy}
\affiliation{INFN Sezione di Milano - Bicocca, Piazza della Scienza 3, I-20126 Milano, Italy}

\author{M.~Consonni}
\affiliation{Dipartimento di Fisica, Università di Milano - Bicocca, Piazza della Scienza 3, I-20126 Milano, Italy}
\affiliation{INFN Sezione di Milano - Bicocca, Piazza della Scienza 3, I-20126 Milano, Italy}

\author{G.~Croci}
\affiliation{Dipartimento di Fisica, Università di Milano - Bicocca, Piazza della Scienza 3, I-20126 Milano, Italy}
\affiliation{INFN Sezione di Milano - Bicocca, Piazza della Scienza 3, I-20126 Milano, Italy}

\author{I.~Dafinei}
\affiliation{INFN Sezione di Roma, P.le Aldo Moro 2, I-00185 Roma, Italy}

\author{F.A.~Danevich}
\affiliation{Institute for Nuclear Research of NASU, 03028 Kyiv, Ukraine}
\affiliation{Institute of Experimental and Applied Physics, Czech Technical University in Prague, 11000 Prague, Czech Republic}

\author{C.~de~Vecchi}
\affiliation{INFN Sezione di Pavia, Via Bassi 6, I-27100 Pavia, Italy}

\author{D.~Di~Martino}
\affiliation{Dipartimento di Fisica, Università di Milano - Bicocca, Piazza della Scienza 3, I-20126 Milano, Italy}
\affiliation{INFN Sezione di Milano - Bicocca, Piazza della Scienza 3, I-20126 Milano, Italy}

\author{R.~Elleboro}
\affiliation{INFN Laboratori Nazionali del Gran Sasso, Via G. Acitelli 22, I-67100 Assergi, Italy}
\affiliation{Dipartimento di Scienze Fisiche e Chimiche, Università degli Studi dell'Aquila, I-67100 L'Aquila, Italy}

\author{N.~Ferreiro Iachellini}
\email{Contact author: nahuel.ferreiroiachellini@unimib.it}
\affiliation{Dipartimento di Fisica, Università di Milano - Bicocca, Piazza della Scienza 3, I-20126 Milano, Italy}
\affiliation{INFN Sezione di Milano - Bicocca, Piazza della Scienza 3, I-20126 Milano, Italy}

\author{F.~Ferroni}
\affiliation{Gran Sasso Science Institute, Viale F. Crispi 7, I-67100 L’Aquila, Italy}
\affiliation{INFN Sezione di Roma, P.le Aldo Moro 2, I-00185 Roma, Italy}

\author{F.~Filippini}
\affiliation{INFN Sezione di Milano - Bicocca, Piazza della Scienza 3, I-20126 Milano, Italy}
\affiliation{DISAT, Università di Milano - Bicocca, Piazza della Scienza 1, I-20126 Milano, Italy}

\author{S.~Ghislandi}
\affiliation{Massachusetts Institute of Technology, Cambridge, MA 02139, USA}


\author{A.~Giachero}
\affiliation{Dipartimento di Fisica, Università di Milano - Bicocca, Piazza della Scienza 3, I-20126 Milano, Italy}
\affiliation{INFN Sezione di Milano - Bicocca, Piazza della Scienza 3, I-20126 Milano, Italy}

\author{L.~Gironi}
\affiliation{Dipartimento di Fisica, Università di Milano - Bicocca, Piazza della Scienza 3, I-20126 Milano, Italy}
\affiliation{INFN Sezione di Milano - Bicocca, Piazza della Scienza 3, I-20126 Milano, Italy}

\author{P.~Gorla}
\affiliation{INFN Laboratori Nazionali del Gran Sasso, Via G. Acitelli 22, I-67100 Assergi, Italy}

\author{C.~Gotti}
\affiliation{Dipartimento di Fisica, Università di Milano - Bicocca, Piazza della Scienza 3, I-20126 Milano, Italy}
\affiliation{INFN Sezione di Milano - Bicocca, Piazza della Scienza 3, I-20126 Milano, Italy}

\author{D.L.~Helis}
\affiliation{INFN Laboratori Nazionali del Gran Sasso, Via G. Acitelli 22, I-67100 Assergi, Italy}

\author{D.V.~Kasperovych}
\affiliation{Institute for Nuclear Research of NASU, 03028 Kyiv, Ukraine}

\author{V.V.~Kobychev}
\affiliation{Institute for Nuclear Research of NASU, 03028 Kyiv, Ukraine}

\author{G.~Marcucci}
\affiliation{Dipartimento di Fisica, Università di Milano - Bicocca, Piazza della Scienza 3, I-20126 Milano, Italy}
\affiliation{INFN Sezione di Milano - Bicocca, Piazza della Scienza 3, I-20126 Milano, Italy}

\author{A.~Melchiorre}
\affiliation{INFN Laboratori Nazionali del Gran Sasso, Via G. Acitelli 22, I-67100 Assergi, Italy}
\affiliation{Dipartimento di Scienze Fisiche e Chimiche, Università degli Studi dell'Aquila, I-67100 L'Aquila, Italy}

\author{A.~Menegolli}
\affiliation{INFN Sezione di Pavia, Via Bassi 6, I-27100 Pavia, Italy}
\affiliation{Dipartimento di Fisica, Università di Pavia, Via Bassi 6, I-27100 Pavia, Italy}

\author{S.~Nisi}
\affiliation{INFN Laboratori Nazionali del Gran Sasso, Via G. Acitelli 22, I-67100 Assergi, Italy}

\author{M.~Musa}
\affiliation{Dipartimento di Scienze della Terra e dell'Ambiente, Università di Pavia, Via Ferrata 7, I-27100 Pavia, Italy}

\author{L.~Pagnanini}
\affiliation{Gran Sasso Science Institute, Viale F. Crispi 7, I-67100 L’Aquila, Italy}
\affiliation{INFN Laboratori Nazionali del Gran Sasso, Via G. Acitelli 22, I-67100 Assergi, Italy}

\author{L.~Pattavina}
\affiliation{Dipartimento di Fisica, Università di Milano - Bicocca, Piazza della Scienza 3, I-20126 Milano, Italy}
\affiliation{INFN Sezione di Milano - Bicocca, Piazza della Scienza 3, I-20126 Milano, Italy}

\author{G.~Pessina}
\affiliation{INFN Sezione di Milano - Bicocca, Piazza della Scienza 3, I-20126 Milano, Italy}

\author{S.~Pirro}
\affiliation{INFN Laboratori Nazionali del Gran Sasso, Via G. Acitelli 22, I-67100 Assergi, Italy}

\author{S.~Pozzi}
\affiliation{Dipartimento di Fisica, Università di Milano - Bicocca, Piazza della Scienza 3, I-20126 Milano, Italy}
\affiliation{INFN Sezione di Milano - Bicocca, Piazza della Scienza 3, I-20126 Milano, Italy}

\author{M.C.~Prata}
\affiliation{INFN Sezione di Pavia, Via Bassi 6, I-27100 Pavia, Italy}

\author{A.~Puiu}
\affiliation{INFN Laboratori Nazionali del Gran Sasso, Via G. Acitelli 22, I-67100 Assergi, Italy}

\author{S.~Quitadamo}
\affiliation{Dipartimento di Fisica, Università di Milano - Bicocca, Piazza della Scienza 3, I-20126 Milano, Italy}
\affiliation{INFN Sezione di Milano - Bicocca, Piazza della Scienza 3, I-20126 Milano, Italy}

\author{M.P.~Riccardi}
\affiliation{Dipartimento di Scienze della Terra e dell'Ambiente, Università di Pavia, Via Ferrata 7, I-27100 Pavia, Italy}

\author{M.~Rossella}
\affiliation{INFN Sezione di Pavia, Via Bassi 6, I-27100 Pavia, Italy}

\author{R.~Rossini}
\affiliation{Dipartimento di Fisica, Università di Pavia, Via Bassi 6, I-27100 Pavia, Italy}
\affiliation{INFN Sezione di Pavia, Via Bassi 6, I-27100 Pavia, Italy}

\author{E.~Sala}
\affiliation{INFN Sezione di Milano - Bicocca, Piazza della Scienza 3, I-20126 Milano, Italy}
\affiliation{Center for Underground Physics, Institute for Basic Science, 34126 Daejeon, Korea}

\author{F.~Saliu}
\affiliation{INFN Sezione di Milano - Bicocca, Piazza della Scienza 3, I-20126 Milano, Italy}
\affiliation{DISAT, Università di Milano - Bicocca, Piazza della Scienza 1, I-20126 Milano, Italy}

\author{A.~Salvini}
\affiliation{Laboratorio Energia Nucleare Applicata, Via Aselli 41, I-27100 Pavia, Italy}

\author{V.I.~Tretyak}
\affiliation{Institute for Nuclear Research of NASU, 03028 Kyiv, Ukraine}
\affiliation{Institute of Experimental and Applied Physics, Czech Technical University in Prague, 11000 Prague, Czech Republic}

\author{L.~Trombetta}
\affiliation{Dipartimento di Fisica, Università di Milano - Bicocca, Piazza della Scienza 3, I-20126 Milano, Italy}
\affiliation{INFN Sezione di Milano - Bicocca, Piazza della Scienza 3, I-20126 Milano, Italy}

\author{D.~Trotta}
\affiliation{Dipartimento di Fisica, Università di Milano - Bicocca, Piazza della Scienza 3, I-20126 Milano, Italy}
\affiliation{INFN Sezione di Milano - Bicocca, Piazza della Scienza 3, I-20126 Milano, Italy}

\author{H.~Yuan}
\affiliation{Shanghai Institute of Ceramics, CAS, 1295 Dingxi Road, Shanghai 200050, P.R. China}

\collaboration{The RES-NOVA Collaboration}\email{Contact collaboration: res-nova@unimib.it}
\author{J. Luengas}\affiliation{\udel}
\author{H. Ramani}\affiliation{\udelbartol}
\noaffiliation

\date{\today}
\begin{abstract}
Probing inelastic dark matter at large mass splittings requires heavy target nuclei, an extended recoil-energy range, and the high-velocity tail of the dark-matter distribution. We exploit these features with the RES-NOVA prototype detector, featuring a $\mathrm{PbWO}_4$ cryogenic calorimeter, produced from archaeological Pb and operated at the deep-underground laboratory of Gran Sasso of INFN (Italy), analyzing a $32.4\,\mathrm{g\,day}$ exposure over $2.5\,\mathrm{keV}$--$1\,\mathrm{MeV}$ under both the Standard Halo Model (SHM) and a Large Magellanic Cloud (LMC)--motivated velocity distribution. We extend direct-detection constraints beyond the $330\,\mathrm{keV}$ reach of established technologies (e.g. Xe-based TPCs), probing splittings up to $510 \ (780)\,\mathrm{keV}$ in the SHM (LMC) benchmark, while future exposures will probe new regions of the parameter space.
\end{abstract}

\maketitle
\textit{Introduction.}\textemdash
The nature of dark matter remains one of the central open questions in particle physics and cosmology. Weakly interacting massive particles (WIMPs) provide a well-motivated class of candidates~\cite{ARBEY2021103865}, but conventional nuclear-recoil searches have placed stringent constraints on elastic dark matter scattering~\cite{PhysRevD.31.3059}. One alternative is provided by inelastic dark matter (iDM) ~\cite{Hall_1998,Smith_2001}, where the dark sector contains two nearly degenerate states, $\chi$ and $\chi'$, separated by a mass splitting $\delta$. In this context, up-scattering off a nucleus is allowed only when the incoming particle carries enough energy to excite the heavier state. As a result, sufficiently large mass splittings can suppress conventional nuclear-recoil signals and allow iDM to evade stringent elastic-search constraints~\cite{Smith_2001,Bramante_2016}. This has motivated a broad range of complementary detection strategies, including luminous and two-step searches~\cite{Feldstein_2010,Pospelov_2014}, as well as searches using metastable nuclear isomers to probe generic inelastic splittings~\cite{Pospelov_2020,Lehnert_2020,Arnquist_2023,alves2023darkmatterconstraintsisomeric,Belli:2025rlx}. Complex electroweak multiplets with nonzero hypercharge ($Y$) provide a particularly well-motivated realization of this scenario \cite{Bottaro_2022}, including the Higgsino, one of the last supersymmetric (SUSY) electroweak WIMPs \cite{Krall_2018}, making the development of new probes of the inelastic frontier especially compelling. 

The maximum mass splitting accessible in inelastic DM--nucleus scattering scales with the DM--nucleus reduced mass and the maximum DM velocity in the detector frame~\cite{PhysRevD.111.055030}. Heavy targets such as lead and enhanced high-velocity DM populations therefore provide a direct kinematic advantage over conventional xenon-based searches, pushing the inelastic frontier to larger mass splittings~\cite{Pospelov_2014, Eby_2019,Graham:2024syw,Graham:2026ivn}. At the same time, large splittings preferentially populate high nuclear-recoil energies~\cite{Bramante_2016}, making the restricted low-energy analysis windows of conventional direct-detection experiments an additional limitation on their reach.

RES-NOVA is particularly well positioned to overcome both limitations. Its archaeological-Pb-based PbWO$_4$ cryogenic calorimeters~\cite{RES-NOVACollaboration:2025stq}, originally developed for coherent elastic neutrino--nucleus scattering~\cite{Pattavina:2020cqc,RES-NOVA:2021gqp,RN_NSI}, combine a heavy Pb-rich target, sensitivity to high recoil energies, and ultra-low radioactive backgrounds. The sensitivity of PbWO$_4$ to iDM was first explored in Ref.~\cite{Song:2021yar} through a recast of alpha-decay search data~\cite{Beeman:2012wz}. However, that interpretation relied on astrophysical assumptions that are not directly comparable to the recommended Standard Halo Model (SHM) parameterization adopted here~\cite{Baxter:2021pqo}.

In this Letter, we use a RES-NOVA prototype detector as a Pb-based nuclear-recoil probe of inelastic dark matter and project the reach of future exposures. We compute the expected recoil spectrum in PbWO$_4$. A dedicated neutron-calibration campaign fully characterized the detector response to nuclear recoils and enabled the independent identification of the region of interest, which is remarkably background-free above 600~keV. We present limits in the $(\sigma_{\rm SI},\delta)$ plane for $m_\chi=1.1~\mathrm{TeV}$, motivated by the Higgsino benchmark, under both the Standard Halo Model~\cite{Baxter:2021pqo,Drukier:1986tm} (SHM) and a Large-Magellanic-Cloud-motivated velocity distribution~\cite{Smith-Orlik_2023} (LMC). Beyond the Higgsino benchmark, we interpret the projected sensitivity for complex electroweak $n$-plets, presenting the reach in the $(M,\delta)$ plane together with the thermal-relic masses of viable multiplets.
In this manuscript we present the results obtained with a proof-of-principle detector, together with the current exclusion limits and the projected reach of the proposed technology.

\textit{iDM--nucleus scattering.}\textemdash
We consider a dark matter particle $\chi$ with an excited state $\chi'$ separated by a positive mass splitting $\delta$. The interaction with ordinary matter is taken to be off diagonal, so that the leading nuclear process is upscattering, $\chi A\rightarrow \chi' A$. This structure is realized, for example, by electroweak multiplet dark matter with nonzero hypercharge \cite{Bottaro_2022}. In the Higgsino limit \cite{Krall_2018}, electroweak symmetry breaking splits a neutral Dirac fermion into two Majorana states, and the $Z$-boson coupling becomes off diagonal in the mass basis. The corresponding per-nucleon cross section is large on direct-detection scales \cite{PhysRevD.31.3059}, but the process is entirely forbidden when the splitting exceeds the kinetic energy available in the incoming dark matter population.

For a nuclear recoil energy $E_R$, the minimum incoming speed required for upscattering is
\begin{equation}\label{eq:vmin}
v_{\rm min}(E_R)=
\frac{1}{\sqrt{2m_AE_R}}
\left(
\frac{m_AE_R}{\mu_A}+\delta
\right),
\end{equation}
where $m_A$ is the target nucleus mass and $\mu_A$ is the dark matter--nucleus reduced mass~\cite{PhysRevD.111.055030}. The absolute minimum speed required for any up-scattering, obtained by minimizing Eq.~\ref{eq:vmin} over $E_R$, is:
\begin{equation}
v_{\rm min}^{\rm global}=\sqrt{\frac{2\delta}{\mu_A}},
\qquad
\delta_{\rm max,A}=\frac{1}{2}\mu_A v_{\rm max}^2 .
\label{eq:deltamax}
\end{equation}
For spin-independent scattering normalized to the per-nucleon cross section $\sigma_{\rm SI}$, the differential rate per target nucleus can be written as~\cite{PhysRevD.111.055030}
\begin{equation}
\frac{d\Gamma_A}{dE_R}=
\frac{\rho_\chi}{m_\chi}
\frac{m_A\sigma_{\rm SI}}{2\mu_n^2}
A^2 F_A^2(E_R)
\eta\left(v_{\rm min}(E_R)\right),
\end{equation}
where $\rho_\chi$ is the local dark matter density, $\mu_n$ is the dark matter--nucleon reduced mass, $F_A(E_R)$ is the nuclear form factor \cite{Eby_2019}, and
\begin{equation}
\eta(v_{\rm min})=
\int_{v>v_{\rm min}} d^3v,
\frac{f_{\rm lab}(\mathbf v)}{v}
\end{equation}
is the mean inverse speed in the laboratory frame~\cite{Smith-Orlik_2023}. The total PbWO$_4$ recoil spectrum is obtained by summing over each element contributions with their corresponding abundances and detector response.

Since the predicted limits are sensitive to the velocity distribution
$f_{\rm lab}(\mathbf{v})$~\cite{Herrera_2023}, we evaluate $\eta(v_{\min})$
for both the SHM and the LMC velocity distributions. The latter is a
simulation-based distribution, motivated by the presence of the Large
Magellanic Cloud, which has been shown to increase collisional
direct-detection (CDD) limits~\cite{Smith-Orlik_2023,reynosocordova2024largemagellaniccloudexpanding}
and gives a more optimistic reach at large $\delta$ owing to its enhanced
high-velocity population.

\begin{figure*}[t]
    \centering

    \includegraphics[width=0.48\textwidth]{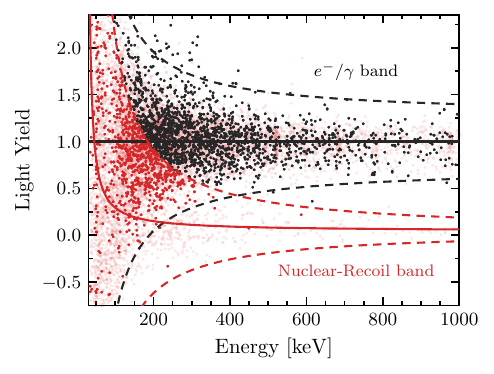}
    \hfill
    \includegraphics[width=0.48\textwidth]{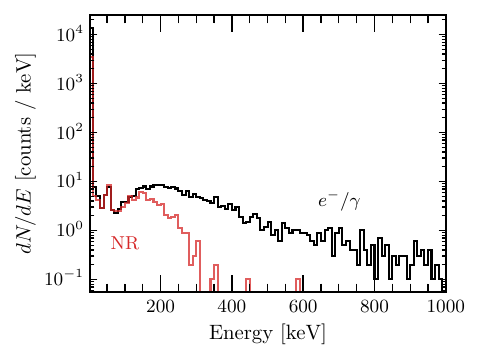}

    \caption{Event discrimination and measured spectra in the RES-NOVA prototype detector.
    (a) Data taken with the detector module in the light-yield--energy plane. Black and pale-red points corresponds to events acquired during background and neutron calibration runs, respectively. Dark-red events refers to background events falling in the nuclear-recoil band. The dashed-lines (solid-lines) indicates the $\pm3\sigma$ bands (central values) for $e^-/\gamma$ and nuclear-recoil interactions. Above $\sim$420~keV, the two populations are fully separated at the 3$\sigma$ level.
    (b) Measured energy spectra for all events and events selected within the nuclear-recoil band.}
    \label{fig:lightyield_background}
\end{figure*}

\textit{Experimental set-up.}\textemdash
The measurement was performed with the RES-NOVA prototype detector described in Ref.~\cite{alloni2026probingdarkmatterinteractions}, its target is a $0.7\times0.7\times4~\mathrm{cm^3}$, $13~\mathrm{g}$ PbWO$_4$ crystal grown from archaeological Pb and cut from the same boule used to characterize the intrinsic radiopurity of this material~\cite{kg-scale}. The crystal was held in position by PTFE clamps inside an OFHC copper frame and operated as a cryogenic calorimeter, with the heat signal read out by a neutron-transmutation-doped (NTD) Ge thermistor glued to its surface. The detector was mounted in the Ieti dry dilution refrigerator~\cite{IETI}, a facility optimized for mechanical stability and vibrational isolation and reaching a base temperature below $7~\mathrm{mK}$, installed in Hall~C of the deep-underground laboratory of Gran Sasso of INFN (Italy), under an overburden of $3600~\mathrm{m\,w.e.}$~\cite{G.Bellini_2012}. A passive shield made of $10~\mathrm{cm}$ of commercial Pb on the sides and $6~\mathrm{cm}$ on top, with additional $5~\mathrm{cm}$ Cu layer, surrounds the detector. We refer the reader to Ref.~\cite{alloni2026probingdarkmatterinteractions} for a complete description of the cryogenic infrastructure, shielding, and low-noise readout electronics.

For the present analysis the calorimetric readout (heat) is combined with a scintillation-light measurement. PbWO$_4$ is a scintillator, and a small fraction of the deposited energy is emitted as scintillation light at cryogenic temperature~\cite{Beeman:2011kv}. The crystal is faced by a cryogenic light detector (LD), consisting of a high-purity Ge disk operated as a bolometer and instrumented with its own NTD-Ge thermistor, of the same type developed and operated by the CUPID-0 experiment~\cite{Azzolini:2018tum, Azzolini:2019tta}. Since the crystal light yield depends on the particle interaction type, electron and $\gamma$ interactions producing more light per unit deposited energy than the strongly quenched $\alpha$ particles and nuclear recoils, the time-coincident heat and light signals provide an event-by-event particle-identification observable. We exploit it to identify and constrain the $\alpha$-induced and nuclear recoil background that populates the spectrum in the region of interest, allowing the statistical treatment described below.
The light detector was not optimized for this specific measurement, as its design and mounting scheme were not tailored to detect low-energy scintillation light signals.

\begin{figure*}[t]
    \centering
   \includegraphics[width=0.48\textwidth]{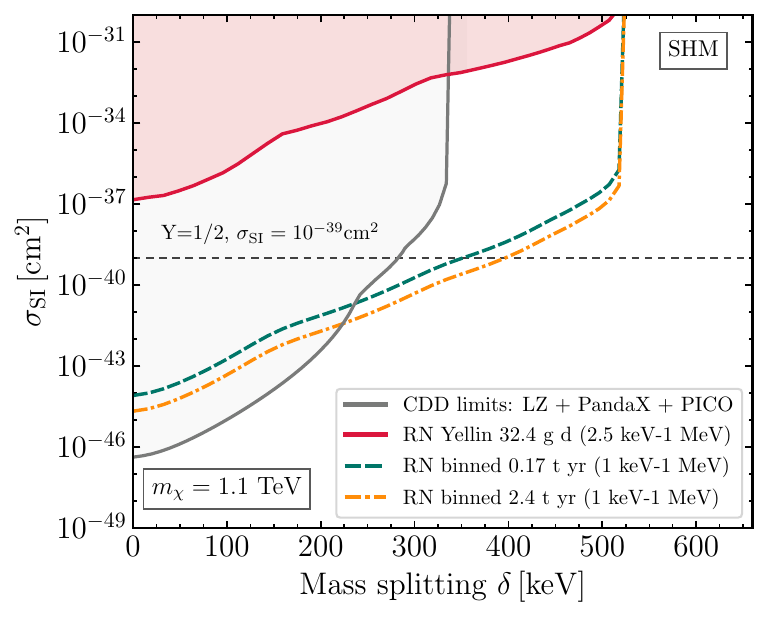}
\hfill
\includegraphics[width=0.48\textwidth]{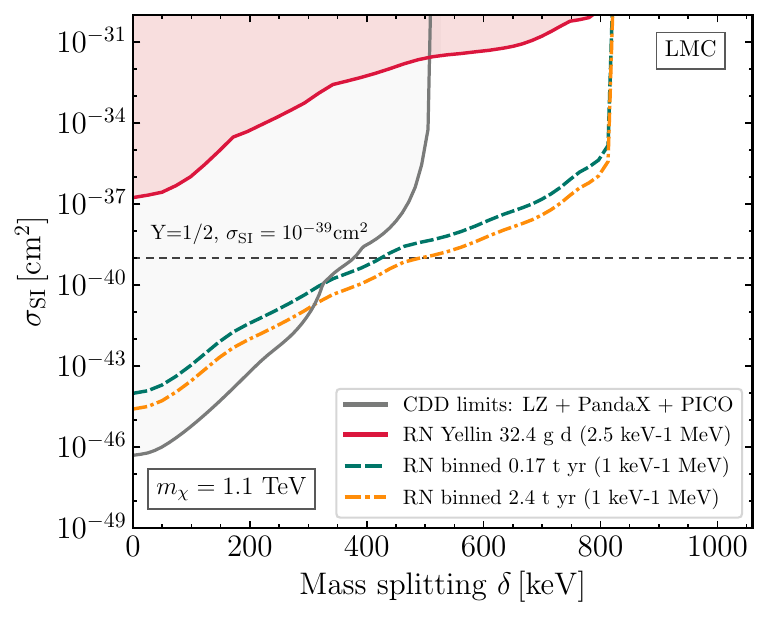}
    \caption{Current sensitivity and the projected $90\%$ C.L. upper limit from RES--NOVA to inelastic dark matter for $m_\chi = 1.1~\mathrm{TeV}$, the left and right panel assume the SHM and LMC velocity distributions, respectively. The red curve shows observed limits obtained with Yellin's optimum-interval method using the $2.5$--$1000~\mathrm{keV}$ nuclear-recoil dataset \cite{alloni2026probingdarkmatterinteractions}. Green and orange curves show projected sensitivities for exposures of $0.17$ and $2.4$ tonne-years, respectively, based on the simulated expected RES--NOVA background~\cite{RES-NOVACollaboration:2025stq}. The gray curve shows the combined current collisional direct-detection (CDD) constraint from LZ, PandaX-4T, and PICO~\cite{Aalbers_2024,PhysRevLett.127.261802,adams2023searchinelasticdarkmatternucleus}. Horizontal dashed line indicate inelastic scattering cross section, $\sigma_{\rm SI}= 10^{-39}~\mathrm{cm}^2$, for complex electroweak multiplets with hypercharge $Y=1/2$~\cite{PhysRevD.111.055030}.
    }
    \label{fig:Results}
\end{figure*}

\textit{Data analysis.}\textemdash 
The continuous data stream produced by the readout chain of the two channels, corresponding to a total exposure of 32.4~g$\cdot$d, was processed through a trigger-less analysis pipeline specifically developed for RES-NOVA. The raw data stream of the calorimetric channel is replicated into two parallel branches: a filtered branch utilized for precise event time-tagging via a low-pass filter, and an unfiltered branch used for pulse-shape characterization.

For each identified event, the pulse amplitude is reconstructed using two independent estimators: an Optimum Filter (OF)~\cite{Gatti:1986cw}, which maximizes the signal-to-noise ratio ($\mathrm{S/N}$) based on the average signal template and noise power spectral density, and a Maximum-Likelihood Estimation (MLE) fit. Non-physical or poorly reconstructed events (e.g., pile-up, baseline instabilities) are rejected by applying a quality cut on the relative difference between the two estimators: $|A_{\mathrm{OF}}-A_{\mathrm{MLE}}|/A_{\mathrm{OF}} < \epsilon_A$. Adopting a conservative tolerance of $\epsilon_A = 10\%$, we achieve a baseline resolution of $\sigma = 234$~eV, ensuring a linear detector response down to the analysis energy threshold of 2.5~keV ($\mathrm{S/N} \sim 1$).

The MLE procedure performs a simultaneous fit on both heat and light channels by minimizing their combined $\chi^2$, with the pulse onset times strictly constrained to account for event coincidence. This joint minimization yields distinct amplitudes for each channel. 

The absolute energy scale of the calorimetric channel was anchored to the 2615~keV $\gamma$ line of $^{208}$Tl from a calibration measurement, corresponding to a signal amplitude of 112~$\mu$V/keV. Low-energy calibration lines, such as 46~keV $\gamma$ emission of $^{210}$Pb, were used to control the reliability of the absolute energy scale over a broad energy range.

No quality selection cuts were applied to the light channel to prevent the introduction of additional systematic uncertainties, given its sub-optimal performance.

\textit{Event Discrimination and Acceptance Region.}\textemdash
The light yield is defined as the ratio of the light signal to the calorimetric energy signal, and exhibits a two-band structure, as shown in Fig.~\ref{fig:lightyield_background}. A rotated two-dimensional Gaussian likelihood fit is used to determine the centroids and widths of the electron-recoil/$\gamma$ band ($e^{-}/\gamma$, black) and the nuclear-recoil band (NR, pale red). We use an energy-independent parametrization for the light channel resolution. The corresponding fitted centroids are $\mu_{\rm ER}=1.00$ and $\mu_{\rm NR}=0.073$, where the light yield has been normalized to the $e^-/\gamma$ band. For the NR band, the inferred energy-dependent correction to the resolution is only of order $10^{-4}$ relative to the leading energy-independent contribution, and it does not modify the set of events accepted in the signal region. Nuclear recoils exhibit a reduced light output, dictated by their quenching factors. The NR band was populated by exposing the detector module to an AmBe neutron source, which also induces $\gamma$ rays; as a result, both the electron-recoil/$\gamma$ band and the nuclear-recoil band are populated in the same calibration run.

The dark matter acceptance region is defined a priori in the energy interval $2.5~{\rm keV} \leq E \leq 1~{\rm MeV}$. In the light-yield plane, it is taken to be the region within $\pm\ 3 \sigma_{\rm NR}$ of the fitted NR-band centroid $\mu_{\rm NR}$. Events falling within this acceptance region, shown as dark red points in Fig.~\ref{fig:lightyield_background}, are conservatively treated as potential signal candidates in the exclusion-limit calculation.

\begin{figure*}[t]
    \centering
    
       \includegraphics[width=0.48\textwidth]{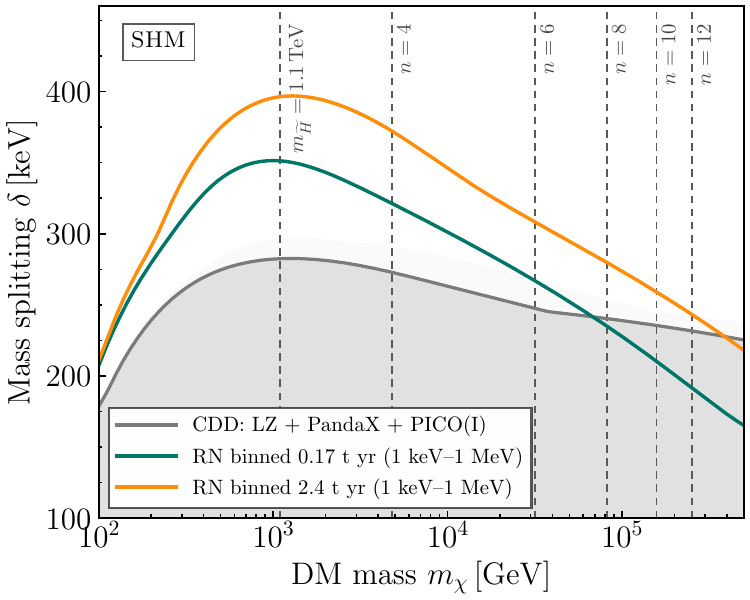}
        \hfill
        \includegraphics[width=0.48\textwidth]{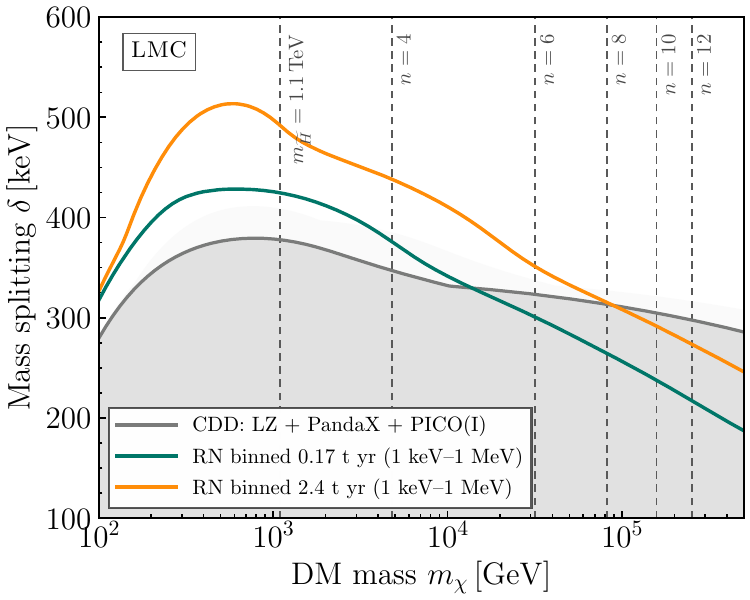}
    \caption{Current and projected reach in the mass splitting $\delta$ as a function of the dark matter mass $m_\chi$, assuming an inelastic spin-independent cross section $\sigma_{\rm SI}=10^{-39}\,{\rm cm}^2$. 
    The left and right panels show the results for the Standard Halo Model (SHM) and the Large Magellanic Cloud (LMC) velocity distribution, respectively. 
    The gray curve shows the current CDD constraints, while the green and orange curves show the projected RES-NOVA binned sensitivities for exposures of $0.17$ and $2.4$ tonne-years, respectively. 
    The vertical dashed lines indicate the thermal relic masses of the last complex $n$-plets with hypercharge $Y=1/2$.}
    \label{fig:delta_mchi_map}
\end{figure*}

\textit{Statistical treatment}\textemdash 
Limits are derived using Yellin's optimum interval method~\cite{Yellin_2002} and a binned Poisson likelihood analysis~\cite{Cowan_2011}. The former is applied to the observed data, where we did not perform any background model reconstruction, and the latter is used to directly include the background modeling into the statistical inference. Specifically, the limit from the 32.4~g$\cdot$d exposure is obtained by applying Yellin's \textit{optimum interval method} to the observed NR-band events over the 2.5~keV--1~MeV analysis range shown in Fig.~\ref{fig:Results}, whereas the future RES--NOVA projections are derived using the binned likelihood analysis with the simulated background model of Ref.~\cite{RES-NOVACollaboration:2025stq}.

\textit{Results}\textemdash
Motivated by the search for new probes of the last viable electroweak WIMPs~\cite{Bottaro_2022}, we first adopt the thermal Higgsino as a benchmark. It corresponds to the $n=2$ fermionic $Y=1/2$ complex multiplet, one of the last viable SUSY electroweak WIMP candidates, with $m_\chi\simeq1.1~\mathrm{TeV}$. In this context, RES--NOVA extends the collisional direct-detection limits for inelastic dark matter to large mass splittings through the combination of its heavy Pb target and high-energy recoil acceptance. The heavy Pb target ($A\simeq207$) increases the maximum kinematically accessible splitting to $\delta_{\rm Pb}^{\rm max}\simeq530~(830)~\mathrm{keV}$ for the SHM (LMC), compared with $\delta_{\rm Xe}^{\rm max}\simeq350~(550)~\mathrm{keV}$ for Xe ($A\simeq131$). The larger values obtained for the LMC provide a clear example of the impact of the halo velocity distribution on the reach, since its enhanced high-velocity population allows larger mass splittings to be accessed. In addition, the high-energy acceptance of the analysis includes nuclear recoils with calibrated efficiency up to $1~\mathrm{MeV}$. At large $\delta$, the recoil energy that minimizes the required incoming velocity satisfies $E_R\simeq\delta$, shifting the signal toward recoil energies of several hundred keV. Such events fall outside the recoil-energy windows used by current CDD searches, which extend to $75~\mathrm{keV}$ for LZ~\cite{Aalbers_2024} and $\mathcal{O}(100~\mathrm{keV})$ for PandaX-4T and PICO~\cite{PhysRevLett.127.261802,adams2023searchinelasticdarkmatternucleus}, while RES--NOVA retains sensitivity to this high-$\delta$ signal.

These advantages are directly reflected in Fig.~\ref{fig:Results}, which shows the corresponding $90\%$ C.L. upper limits in the $(\sigma_{\rm SI}$, $\delta)$ plane and clearly illustrates the extended reach of RES--NOVA over current searches. The gray curve corresponds to the current CDD limits, which rule out values up to $\delta_{\rm CDD}^{\rm max}\simeq330~(500)~\mathrm{keV}$ for the SHM (LMC) at cross sections above approximately $\sigma_{\rm SI}=10^{-36}~\mathrm{cm}^2$. The solid red curve corresponds to the current $32.4~\mathrm{g\cdot d}$ prototype exposure~\cite{alloni2026probingdarkmatterinteractions}, using events detected within the NR band. Already with this exposure, RES--NOVA extends the excluded region up to $\delta_{\rm RN}^{\rm max}\simeq510~(780)~\mathrm{keV}$ for the SHM (LMC), at cross sections above approximately $\sigma_{\rm SI}=10^{-32}~\mathrm{cm}^2$. Thus, RES--NOVA already probes previously unconstrained parameter space over $\delta\simeq330$--$510~\mathrm{keV}$ in the SHM and $\delta\simeq500$--$780~\mathrm{keV}$ in the LMC model, demonstrating a clear extension beyond current CDD searches with existing data, whereas at smaller splittings and cross sections the current CDD limits remain stronger by several orders of magnitude.

Future RES--NOVA projections approach the kinematic limit of Pb for cross sections above approximately $\sigma_{\rm SI}=10^{-36}~\mathrm{cm}^2$, allowing the high-$\delta$ region discussed above to be probed. Increasing the exposure primarily improves the reach toward smaller cross sections. With exposures of $0.17$ and $2.4$ tonne$\cdot$years~\cite{Pattavina:2020cqc} (green and orange curves), RES--NOVA reaches the characteristic complex-$n$-plet cross section for $Y=1/2$, becoming sensitive to some of the last surviving WIMP candidates, including the high-$\delta$ parameter space beyond the reach of Xe experiments.

Beyond the Higgsino benchmark, a complementary view of the projected reach is shown in Fig.~\ref{fig:delta_mchi_map}, where we fix the inelastic spin-independent cross section to $\sigma_{\rm SI}=10^{-39}~\mathrm{cm^2}$ and show sensitivity to mass splitting $\delta$ as a function of the dark matter mass. Extending the interpretation to the full set of viable fermionic $Y=1/2$ complex electroweak multiplets, the vertical dashed lines indicate their standard thermal freeze-out masses for $n=2,4,6,8,10,$ and $12$~\cite{Bottaro_2022}. These masses reproduce the observed dark matter abundance through thermal freeze-out in a standard cosmological history. We nevertheless display the sensitivity over a continuous range of $m_\chi$, since nonstandard cosmological histories or additional production mechanisms may modify the relation between the dark matter mass and the observed relic abundance.

The mass dependence can be understood from the interplay between kinematics and rate scaling. At low dark matter masses, the reach is reduced because the dark matter--nucleus reduced mass is smaller, which lowers the available inelastic kinetic energy and suppresses the fraction of the dark matter population capable of producing the scattering. As $m_\chi$ increases, $\mu_A$ approaches $m_A$, allowing larger splittings to be probed until the kinematic gain begins to saturate in the heavy-DM limit. At still larger masses, the overall rate decreases as $\rho_\chi/m_\chi$ which translates to a drop in reach in splitting at a fixed cross-section.

At the Higgsino thermal mass, the 0.17 (2.4)~tonne$\cdot$years RES--NOVA projections reach mass splittings of approximately $\delta\simeq350~(400)~\mathrm{keV}$ in the SHM and $\delta\simeq430~(490)~\mathrm{keV}$ in the LMC, compared with current CDD reaches of approximately $280$ and $380~\mathrm{keV}$, respectively. Beyond the Higgsino benchmark, the 2.4~tonne$\cdot$years projection exceeds the current CDD reach at all thermal $Y=1/2$ multiplet masses shown in the SHM, from $n=2$ to $n=12$, whereas in the LMC it remains clearly superior through $n=6$, becomes comparable near $n=8$, and falls below the current reach for $n=10$ and $12$. The 0.17~tonne$\cdot$years exposure retains an advantage through approximately $n=6$ in the SHM and $n=4$ in the LMC. As seen in Fig.~\ref{fig:delta_mchi_map}, RES--NOVA has discovery potential all the way upto the 100~TeV unitarity bound for electroweak WIMPs.

In summary, by combining a heavy Pb target with sensitivity to MeV-scale nuclear recoils, RES--NOVA opens a unique window onto inelastic dark matter with large mass splittings. Existing prototype data already exclude previously unconstrained parameter space beyond the reach of current dark matter searches, while future exposures extend the sensitivity to substantially smaller cross sections and probe the thermal-relic masses of the last viable complex electroweak WIMPs.

\textit{Acknowledgments}\textemdash
The RES--NOVA collaboration thanks the directors and staff of the Laboratori Nazionali del Gran Sasso for their support, as well as the technical staff involved in the experiment. This work received funding from the European Union’s Horizon Europe programme through the ERC grant ERC--101087295 RES--NOVA. Additional financial support was provided under the National Recovery and Resilience Plan (NRRP) by the Italian Ministry of University and Research (MUR), funded by the European Union – NextGenerationEU, within the project “Advanced techniques for a next--generation isotopically enriched cryogenic dark-matter experiment” (2022L2AXP2). S.~Ghislandi acknowledges support from the U.S. Department of Energy (DOE) Grant No. [DE-SC0011091]. HR and JL are supported by NSF Grant No. PHY-2515007, The University of Delaware Research Foundation and the John Templeton Foundation Award No. 63595. 
\bibliographystyle{apsrev4-1}
\bibliography{reference.bib}

\end{document}